\documentclass[twocolumn]{el-author}
\usepackage{graphicx,amssymb,amsmath}%图片支持
\usepackage{multicol}%正文双栏
\usepackage[noadjust]{cite}%参考文献
\usepackage{setspace}%调整行距
\usepackage{stfloats}
\usepackage{midfloat}
\usepackage[normal]{threeparttable}
\usepackage{flushend,cuted}
\usepackage{cuted}
\usepackage{cases,subeqnarray}
\usepackage{bm,multirow,bigstrut}
\usepackage{textcomp}
\usepackage{latexsym,bm}
\usepackage{booktabs,changebar}
\usepackage{xcolor}
\usepackage{mathtools}
\newcommand{\ua}{\uparrow}
\newcommand{\nc}{\newcommand}
\nc{\da}{\downarrow} \nc{\hc}{\hat{c}} \nc{\hS}{\hat{S}}
\nc{\bra}{\langle} \nc{\ket}{\rangle} \nc{\eq}{equation (\ref}
\nc{\h}{\hat} \nc{\hT}{\h{T}}\nc{\be}{\begin{eqnarray}}
\nc{\ee}{\end{eqnarray}}\nc{\rd}{\textrm{d}}\nc{\e}{eqnarray}\nc{\hR}{\hat{R}}\nc{\Tr}{\mathrm{Tr}}
\nc{\tS}{\tilde{S}}\nc{\tr}{\mathrm{tr}}\nc{\8}{\infty}\nc{\lgs}{\bra\ua,\phi|}\nc{\rgs}{|\ua,\phi\ket}
\nc{\hU}{\hat{U}}\nc{\lfs}{\bra\phi|}\nc{\rfs}{|\phi\ket}\nc{\hZ}{\hat{Z}}\nc{\hd}{\hat{d}}\nc{\mD}{\mathcal{D}}
\nc{\bd}{\bar{d}}\nc{\bc}{\bar{c}}\nc{\mc}{\mathcal}\nc{\ea}{eqnarray}\nc{\mG}{\mathcal{G}}\nc{\bce}{\begin{center}}
\nc{\ece}{\end{center}}
\usepackage{suffix}

\DeclarePairedDelimiterX\MeijerM[3]{\lparen}{\rparen}%
{\begin{smallmatrix}#1 \\ #2\end{smallmatrix}\delimsize\vert\,#3}

\newcommand\MeijerG[8][]{%
  G^{\,#2,#3}_{#4,#5}\MeijerM[#1]{#6}{#7}{#8}}

\WithSuffix\newcommand\MeijerG*[7]{%
  G^{\,#1,#2}_{#3,#4}\MeijerM*{#5}{#6}{#7}}

\begin{document}
\title{Effective capacity of communication systems over $\kappa$-$\mu$ shadowed fading channels}

%\auhtor{Author 1, Author 2, Author 3, Author 4}

\author{Jiayi~Zhang, Linglong~Dai, Wolfgang H. Gerstacker, and Zhaocheng Wang, \textit{IET Fellow}}

\abstract{The effective capacity of communication systems over generalized $\kappa$-$\mu$ shadowed fading channels is investigated in this letter. A novel and analytical expression for the exact effective capacity is derived in terms of extended generalized bivariate Meijer's-$G$ function. To intuitively reveal the impact of the system and channel parameters on the effective capacity, we also derive closed-form expressions for the effective capacity in the asymptotically high signal-to-noise ratio regime. Our results demonstrate that the effective capacity is a monotonically increasing function of channel fading parameters $\kappa$ and $\mu$ as well as the shadowing parameter $m$, while it decays to zero when the delay constraint $\theta \rightarrow \infty$.}

\maketitle

\section{Introduction}
Shannon's ergodic capacity, which has been extensively studied in the literature, cannot account for the quality of service (QoS) requirements of some emerging real-time applications, such as mobile video telephony, interactive gaming, and multimedia streaming. These applications are highly relevant for the next-generation wireless networks, but they need strict delay guarantees. Therefore, a novel performance metric is required to ensure delay guarantees for such real-time applications. Motivated by this fact, the authors of \cite{wu2003effective} proposed the concept of effective capacity (or effective rate, effective throughput) to incorporate statistical delay QoS guarantees in capacity of wireless communications, where the effective capacity is defined as the maximum constant arrival rate at the transmitter when guaranteed statistical delay constraints can be satisfied.

Recently, the effective capacity analysis of wireless communication systems has attracted significant research interests, either in independent fading channels \cite{matthaiou2012analytic} or correlated fading channels \cite{zhong2012effective,guo2012performance,Zhang2014effective}. However, the presented results only consider small-scale fading environment, such as Rayleigh, Rician, Nakagami-$m$, and $\kappa$-$\mu$ fading scenarios, while practical wireless channels suffer from both small-scale fading and shadowing simultaneously. Up to now, the investigation of the impact of shadowing on the effective capacity is still limited.
On the other hand, the $\kappa$-$\mu$ shadowed fading channel, recently proposed in \cite{paris2014statistical}, is a generalized and composite multipath/shadowing fading model. It is based on the assumption that the scattered waves with identical power and the dominant components are subject to the same Nakagami-$m$ shadowing fluctuation, and it has been proved to be able to include Rayleigh, one-sided Gaussian, Rician, Rician shadowed, Nakagami-$m$, and $\kappa$-$\mu$ fading as special cases. Due to its improved flexibility, the $\kappa$-$\mu$ distribution can accurately model the statistical variations of wireless communication channels \cite{paris2014statistical}. To the best of the authors' knowledge, however, results for the effective capacity of $\kappa$-$\mu$ shadowed fading channels are not available in the literature yet.
%This is because significant mathematical challenges for the effective capacity analysis are incurred when considering the rather intricate statistics of generalized $\kappa$-$\mu$ shadowed fading channels.

To this end, we investigate the effective capacity by deriving the exact expressions for any set of parameters of the $\kappa$-$\mu$ shadowed fading channel by using the extended generalized bivariate Meijer's-$G$ function (EGBMGF) \cite[Eq. (1.1)]{shah1973generalizations}. These expressions are generalized and can be reduced to other fading channels. For example, for the special case of the Rician shadowed fading channel, we derive a novel and closed-form expression for the effective capacity. We also derive high-SNR approximated results to draw intuitive insights into the impact of the system and channel parameters on the effective capacity, which reveal that larger values of small-scale fading and shadowing parameters increase the effective capacity, while more stringent delay constraint decreases the effective capacity.

\section{Effective capacity analysis}
As introduced in \cite{wu2003effective}, we suppose that the data arrives in the buffer at a constant rate, and the service process is stationary. A block fading channel is assumed. Analytically, the effective capacity normalized by the bandwidth can be defined as \cite[Eq. (11)]{gursoy2009analysis}
\begin{equation}\label{eq:effective_rate}
R  =  - \frac{1}{A}{\log _2}\left( {{\tt E}\left\{ {{{\left( {1 + \gamma }\right)}^{ - A}}} \right\}} \right) \;\;\textrm{bit/s/Hz},
\end{equation}
where ${\tt E}\{\cdot\}$ refers to the expectation operator, the random variable $\gamma$ denotes the instantaneous signal-to-noise ratio (SNR), $A \triangleq {\theta TB}/{\textrm{ln}2}$ with the asymptotic decay-rate of the buffer occupancy $\theta$, the block length $T$, and the system bandwidth $B$.
From \eqref{eq:effective_rate}, we can find that the effective capacity coincides with the classical Shannon's ergodic capacity when there is no delay constraint as $\theta  \rightarrow 0$.

For the recently proposed $\kappa$-$\mu$ shadowed fading channels \cite{paris2014statistical}, the probability density function (PDF) of $\gamma$ is given by \cite[Eq. (4)]{paris2014statistical}
\begin{align}\label{eq:kappa_mu_shadowed}
{f_\gamma }\left( \gamma  \right)= &\frac{{{\mu ^\mu }{m^m}{{\left( {1 + \kappa } \right)}^\mu }}}{{\Gamma \left( \mu  \right)\bar \gamma {{\left( {\mu \kappa  + m} \right)}^m}}}{\left( {\frac{\gamma }{{\bar \gamma }}} \right)^{\mu  - 1}}\notag \\
&\times {e^{ - \frac{{\mu \left( {1 + \kappa } \right)\gamma }}{{\bar \gamma }}}}{}_1{F_1}\left( {m,\mu ;\frac{{{\mu ^2}\kappa \left( {1 + \kappa } \right)\gamma }}{({\mu \kappa  + m}){\bar \gamma }}} \right),
\end{align}
where $m$ denotes the shaping parameter of the Nakagami-$m$ RV, $\kappa$ represents the power ratio between the dominant components and the scattered waves, $\mu$ is the number of clusters, and $\bar \gamma$ stands for the average SNR. Moreover, $\Gamma(\cdot)$ represents the Gamma function \cite[Eq. (8.310.1)]{gradshtein2000table}, and ${}_1{F_1}(\cdot)$ is the confluent hypergeometric function \cite[Eq. (9.210.1)]{gradshtein2000table}, respectively. Thus, we can derive the effective capacity in \eqref{eq:effective_rate} by averaging the SNR $\gamma$ with the PDF in \eqref{eq:kappa_mu_shadowed}, i.e.,
\begin{align}\label{eq:effective_rate_kappa}
R =  - \frac{1}{A}{\log _2}\left( {\frac{{{\mu ^\mu }{m^m}{{\left( {1 + \kappa } \right)}^\mu }}}{{\Gamma \left( \mu  \right){{\bar \gamma }^\mu }{{\left( {\mu \kappa  + m} \right)}^m}}}I} \right),
\end{align}
where
\begin{align}\label{eq:effective_rate_I}
I \!=\! \int_0^\infty  {{{\left( {1\! +\! \gamma } \right)}^{ \!- \!A}}{\gamma ^{\mu  \!-\! 1}}{e^{ \!-\! \frac{{\mu \left( {1 \!+\! \kappa } \right)\gamma }}{{\bar \gamma }}}}{}_1{F_1}\left( {m,\mu ;\frac{{{\mu ^2}\kappa \left( {1 \!+\! \kappa } \right)\gamma }}{({\mu \kappa \! +\! m}){\bar \gamma }} } \right)} d\gamma.
\end{align}
Note that the integral $I$ in \eqref{eq:effective_rate_I} is still not easy to be calculated. In order to find an analytical solution to the integral $I$, we express ${{\left( {1\! +\! \gamma } \right)}^{ \!- \!A}}$, ${e^{ \!-\! \frac{{\mu \left( {1 \!+\! \kappa } \right)\gamma }}{{\bar \gamma }}}}$, and ${}_1{F_1}\left( {m,\mu ;\frac{{{\mu ^2}\kappa \left( {1 \!+\! \kappa } \right)\gamma }}{({\mu \kappa \! +\! m}){\bar \gamma }} } \right)$ as Meijer's-$G$ functions with the help of \cite[Eqs. (10-11)]{adamchik1990algorithm}, \cite[Eq. (9.34.8)]{gradshtein2000table} and \cite[Eq. (9.212.2)]{gradshtein2000table}, respectively. Then, by using the integral identity \cite[Eq. (07.34.21.0081.01)]{Wolfram2011function}, we can obtain
\begin{align}\label{eq:effective_rate_I_result}
I&= \frac{{\Gamma \left( \mu  \right)}}{{\Gamma \left( A \right)\Gamma \left( {\mu  - m} \right)}}{\left( {\frac{{m\mu \left( {1 + \kappa } \right)}}{{\left( {\mu \kappa  + m} \right)\bar \gamma }}} \right)^{ - \mu }}\notag \\
& \times G_{1,0;1,1;1,2}^{0,1;1,1;1,1}\left[ \! {\left.  \!{\begin{array}{*{20}{c}}
{1 \!- \!\mu }\\
 -
\end{array}} \!\right| \!\left. \! {\begin{array}{*{20}{c}}
{1 \!-\! A}\\
0
\end{array}}  \!\right| \!\left. \! {\begin{array}{*{20}{c}}
{1 \!+\! m \!-\! \mu }\\
{0,1 \!-\! \mu }
\end{array}}  \!\right| \!\frac{{\left( {\mu \kappa  \!+\! m} \right)\bar \gamma }}{{m\mu \left( {1 \!+\! \kappa } \right)}},\frac{{\mu \kappa }}{m}}  \!\right].
\end{align}
The above solution involves an EGBMGF. Although the EGBMGF is not available in standard mathematical packages, an exact and efficient Mathematica implementation has been provided in \cite[Table II]{ansari2011new}, which is based on double Mellin-Barnes type integrals.
Substituting \eqref{eq:effective_rate_I_result} into \eqref{eq:effective_rate_kappa} and performing some algebraic simplifications, the exact effective capacity of $\kappa$-$\mu$ shadowed fading channels can be finally derived as
\begin{align}\label{eq:effective_rate_result}
&R=  - \frac{1}{A}{\log _2}\Bigg( \frac{1}{{\Gamma \left( A \right)\Gamma \left( {\mu  - m} \right)}}{{\left( {\frac{m}{{\mu \kappa  + m}}} \right)}^{m - \mu }}\notag \\
&\times G_{1,0;1,1;1,2}^{0,1;1,1;1,1}\left[ { \!\left. \! {\begin{array}{*{20}{c}}
{1  \!- \! \mu }\\
 -
\end{array}}  \!\right| \!\left. \! {\begin{array}{*{20}{c}}
{1  \!- \! A}\\
0
\end{array}}  \!\right| \!\left. \! {\begin{array}{*{20}{c}}
{1  \!+ \! m  \!- \! \mu }\\
{0,1  \!- \! \mu }
\end{array}}  \!\right| \!\frac{{\left( {\mu \kappa   \!+ \! m} \right)\bar \gamma }}{{m\mu \left( {1  \!+  \!\kappa } \right)}},\frac{{\mu \kappa }}{m}} \!\right] \Bigg).
\end{align}
Note that the derived effective capacity in \eqref{eq:effective_rate_result} is generalized and can be reduced to other fading channels. For example, since $\kappa$-$\mu$ distribution includes the Rician distribution as a special case by setting $\mu=1$, \eqref{eq:effective_rate_result} reduces to the effective capacity of Rician shadowed fading as
\begin{align}\label{eq:effective_rate_Rician_result}
R=  &- \frac{1}{A}{\log _2}\Bigg( \frac{1}{{\Gamma \left( A \right)\Gamma \left( {1 - m} \right)}}{{\left( {\frac{m}{{\kappa  + m}}} \right)}^{m - 1}}\notag \\
&\times G_{1,0;1,1;1,2}^{0,1;1,1;1,1}\left[ {\left. {\begin{array}{*{20}{c}}
0\\
 -
\end{array}} \right|\left. {\begin{array}{*{20}{c}}
{1 - A}\\
0
\end{array}} \right|\left. {\begin{array}{*{20}{c}}
m\\
{0,0}
\end{array}} \right|\frac{{\left( {\kappa  + m} \right)\bar \gamma }}{{m\left( {1 + \kappa } \right)}},\frac{\kappa }{m}} \right] \Bigg),
\end{align}
where $\kappa$ is now identical to the Rician $K$ factor. 

It should be pointed out that although \eqref{eq:effective_rate_result} is the exact effective capacity of $\kappa$-$\mu$ shadowed fading channels, it is only valid for $m \neq \mu$ due to the Gamma function in the denominator. For the special case of $m = \mu$, the PDF of $\kappa$-$\mu$ shadowed fading in \eqref{eq:kappa_mu_shadowed} can be simplified to
\begin{align}\label{eq:effective_rate_PDF_m}
{f_\gamma }\left( \gamma  \right) = \frac{{{m^m}}}{{\Gamma \left( m \right){{\bar \gamma }^m}}}{e^{ - \frac{{m\gamma }}{{\bar \gamma }}}}{\gamma ^{m - 1}},
\end{align}
where the property of \cite[Eq. (07.20.03.0025.01)]{Wolfram2011function} has been used. Substituting \eqref{eq:effective_rate_PDF_m} into \eqref{eq:effective_rate}, using \cite[Eq. (9)]{matthaiou2012analytic} and Kummer's transformation \cite[Eq. (07.33.17.0007.01)]{Wolfram2011function}, the effective capacity $R_{m=\mu}$ in the case of $m = \mu$ is given by
\begin{align}\label{eq:effective_rate_result_m}
R_{m=\mu} = {\log _2}\left( {\frac{\bar \gamma}{{ m}}} \right) - \frac{1}{A}{\log _2}\left( {U\left( {A;A + 1 - m;\frac{m}{{\bar \gamma }}} \right)} \right),
\end{align}
where $U(\cdot)$ is the Tricomi hypergeometric function \cite[Eq. (07.33.02.0001.01)]{Wolfram2011function}.
%It is worth mentioning that an increase in $m$ tends to decrease the effective capacity, since $U(a,b-n,z)$ is a monotonically decreasing function in $n$.

\section{High-SNR analysis}
Although both \eqref{eq:effective_rate_result} and \eqref{eq:effective_rate_result_m} are exact expressions for the effective capacity, they cannot provide intuitive insights into the impact of system and channel parameters on the system performance. Therefore, by considering the initial effective capacity expression \eqref{eq:effective_rate} and keeping only the dominant term when $\gamma \rightarrow \infty$, we can obtain a high-SNR approximation ${R^\infty }$ of the effective capacity as
\begin{align}\label{eq:effective_rate_high}
{R^\infty } \!=\! {\log _2}\left( {\frac{{\bar \gamma }}{{\mu \left( {1 \!+\! \kappa } \right)}}} \right) \!-\! \frac{1}{A}{\log _2}\left( {\frac{{\Gamma \left( {\mu \! -\! A} \right)}}{{\Gamma \left( \mu  \right)}}{}_2{F_1}\left( {m,A;\mu ; \frac{\!-\!{\mu \kappa }}{m}} \right)} \right),
\end{align}
where ${}_2{F_1}(\cdot)$ is the Gauss hypergeometric function \cite[Eq. (9.100)]{gradshtein2000table}. To obtain \eqref{eq:effective_rate_high}, we have used the integral identity \cite[Eq. (7.522.5)]{gradshtein2000table} and a transformation of ${}_2{F_1}(\cdot)$ \cite[Eq. (9.131.1)]{gradshtein2000table}. It is revealed in \eqref{eq:effective_rate_high} that the effective capacity grows logarithmically with the average SNR $\bar \gamma$ in the high-SNR regime.

Moreover, for the special case of $m=\mu$, the approximated effective capacity ${R_{m=\mu}^\infty }$ in the high-SNR regime can be derived as
\begin{align}\label{eq:effective_rate_high_m}
{R_{m=\mu}^\infty } = {\log _2}\left( {\frac{{\bar \gamma }}{m}} \right) - \frac{1}{A}{\log _2}\left( {\frac{{\Gamma \left( {m - A} \right)}}{{\Gamma \left( m \right)}}} \right),
\end{align}
where the function ${}_2{F_1}(\cdot)$ is represented by elementary functions \cite[Eq. (9.121.1)]{gradshtein2000table}. It is revealed in \eqref{eq:effective_rate_high_m} that the effective capacity is a monotonically increasing function of $\mu$ and $m$. This is anticipated, since larger values of $\mu$ refer to more clusters of multipath components, and larger values of $m$ imply that the shadowing effect decreases.

\section{Numerical results}
\begin{figure}[htbp]
\centering
\includegraphics[scale=0.5]{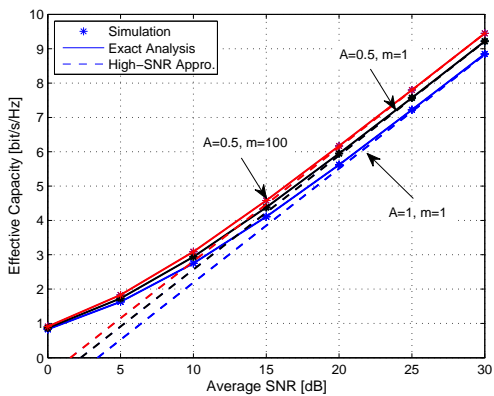}\\
\vspace*{-3mm}\caption{Simulated, exact, and high-SNR approximated effective capacity versus the average SNR $\bar \gamma$ of $\kappa$-$\mu$ shadowed fading channels with different values of $m$ and $A$ ($\kappa = 1$ and $\mu = 2$).}
\label{ER_kappa_mu_shadowed_m_A}
\end{figure}

In Fig. \ref{ER_kappa_mu_shadowed_m_A}, the simulated effective capacity results via Monte-Carlo simulations are compared to the exact and high-SNR approximated analytical expressions provided in \eqref{eq:effective_rate_result} and \eqref{eq:effective_rate_high} for the effective capacity as a function of the average SNR $\bar \gamma$. A precise agreement between the simulated and derived results can be observed, which validates the accuracy of the derived expressions. Moreover, the high-SNR approximations are sufficiently tight and become almost exact when SNR is high, e.g., $\bar \gamma>25$ dB. As anticipated, higher values of $m$ (a weak shadowing condition) tend to result in a higher effective capacity. More importantly, the effective capacity increases when the delay constraint becomes smaller.
\vspace*{-3mm}
%\vspace*{-1mm}
\begin{figure}[htbp]
\centering
\includegraphics[scale=0.5]{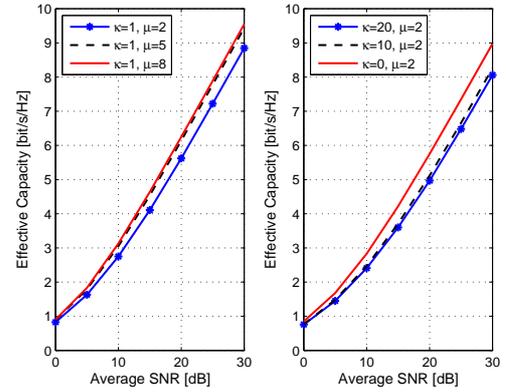}\\
\caption{Exact effective capacity versus the average SNR $\bar \gamma$ of $\kappa$-$\mu$ shadowed fading channels with different values of $\mu$ and $\kappa$ ($m = 1$ and $A = 1$).}
\label{ER_kappa_mu_shadowed}
\vspace*{-4mm}
\end{figure}

Fig. \ref{ER_kappa_mu_shadowed} more deeply investigates the impact of small-scale fading parameters $\kappa$ and $\mu$ on the effective capacity of $\kappa$-$\mu$ shadowed fading channels. As indicated by \eqref{eq:effective_rate_high_m}, a higher $\mu$ yields a higher effective capacity. However, the gap between the corresponding curves decreases as $\mu$ increases, which implies that the effect of $\mu$ becomes less pronounced. On similar grounds, the increase of effective capacity is more pronounced for smaller values of $\kappa$ (less power of dominant components), which reveals that more scattered waves are beneficial for improved effective capacity.

\section{Conclusion}
We have presented a detailed analysis of the effective capacity of $\kappa$-$\mu$ shadowed fading channels in this letter. An exact analytical expression has been obtained. Furthermore, we derived a closed-form expression in the high-SNR regime to gain intuitive insights into the impact of system and channel parameters on the effective capacity. The derived results reveal that a performance gain can be obtained by loosening the delay constraint $\theta$ as well as in propagation environment with larger values of $\mu$ and $m$. Moreover, higher power of dominant components (larger values of $\kappa$) tends to decrease the effective capacity. Additionally, our results can be extended to various small-scale and composite fading channels.

\vspace{-0.5cm}
\vskip3pt
\ack{This work has been supported by the National Key Basic Research Program of China (No. 2013CB329203) and China Postdoctoral Science Foundation (No. 2014M560081).}

\vskip3pt

\noindent J. Zhang, L. Dai (corresponding author) and Z. Wang (\textit{Department of Electronic Engineering as well as Tsinghua
National Laboratory of Information Science and Technology (TNList), Tsinghua University, Beijing 100084, P. R. China})
\vskip3pt

\noindent E-mail: daill@tsinghua.edu.cn

\vskip3pt
\noindent W. H. Gerstacker (\textit{Institute for Digital Communications, University of Erlangen-Nurnberg, D-91058 Erlangen, Germany})

%\bibliographystyle{IEEEtran}
%\bibliography{IEEEabrv,Ref}

\begin{thebibliography}{}

\bibitem{wu2003effective}
Wu D. and Negi R.: `Effective capacity: {A} wireless link model for support of
  quality of service', \emph{IEEE Trans. Wireless Commun.}, 2003, \textbf{2}, (4), pp. 630--643

\bibitem{matthaiou2012analytic}
Matthaiou M., Alexandropoulos G., Ngo H., \textit{et al}: `Analytic framework
  for the effective rate of {MISO} fading channels', \emph{IEEE Trans.
  Commun.}, 2012, \textbf{60}, (6), pp. 1741--1751

\bibitem{zhong2012effective}
Zhong C., Ratnarajah T., Wong K.-K., \textit{et al}: `Effective capacity of
  multiple antenna channels: {C}orrelation and keyhole', \emph{IET Commun.}, 2012, \textbf{6}, (12), pp. 1757--1768

\bibitem{guo2012performance}
Guo X.B., Dong L., and Yang H.: `Performance analysis for effective rate of
  correlated {MISO} fading channels', \emph{Electron. Lett.}, 2012, \textbf{48}, (24), pp. 1564--1565

\bibitem{Zhang2014effective}
Zhang J., Tan Z., Wang H., \textit{et al}: `The effective throughput
  of {MISO} systems over $\kappa$-$\mu$ fading channels', \emph{IEEE Trans.
  Veh. Technol.}, 2014, \textbf{63}, (2), pp. 943--947

\bibitem{paris2014statistical}
Paris J.F.: `Statistical characterization of $\kappa$-$\mu$ shadowed
  fading', \emph{IEEE Trans. Veh. Technol.}, 2014, \textbf{63}, (2), pp. 518--526

\bibitem{gursoy2009analysis}
Gursoy M.C., Qiao D., and Velipasalar S.: `Analysis of energy efficiency in
  fading channels under {QoS} constraints', \emph{IEEE Trans. Wireless
  Commun.}, 2009, \textbf{8}, (8), pp. 4252--4263

\bibitem{gradshtein2000table}
Gradshteyn I.S. and Ryzhik I.M.: \emph{Table of Integrals, Series, and
  Products}, San Diego, CA: Academic Press, 2007.

\bibitem{adamchik1990algorithm}
Adamchik V. and Marichev O.: `The algorithm for calculating integrals of
  hypergeometric type functions and its realization in {REDUCE} system', in
  \emph{Proc. Intern. Conf. Symbolic Algebraic Computation}, 1990, pp.
  212--224.

\bibitem{Wolfram2011function}
Wolfram: `The {W}olfram functions site', {A}vailable:
  \url{http://functions.wolfram.com}.

\bibitem{shah1973generalizations}
Shah M.: \emph{On generalizations of some results and their
  applications}. Seminario Matem{\'a}tico de Barcelona, 1973.

\bibitem{ansari2011new}
Ansari I.S., Al-Ahmadi S., Yilmaz F., \textit{et al}: ``A
  new formula for the {BER} of binary modulations with dual-branch selection
  over generalized-$k$ composite fading channels', \emph{IEEE Trans. Commun.}, 2011, \textbf{59}, (10), pp. 2654--2658


\end{thebibliography}

\end{document}